\documentclass[pra,graphicx, reprint, superscriptaddress]{revtex4-1}

\usepackage{mathtools} 
\usepackage{textcomp}
\usepackage{amssymb}   
\usepackage{siunitx}   
\usepackage{dcolumn}
\usepackage{color}
\usepackage{gensymb}

 \newcommand{\PJ}{$2\,{}^3\!S_1 \rightarrow 2\,{}^3\!P_{\!J}$}
 \newcommand{\Ptwo}{$2\,{}^3\!S_1 \rightarrow 2\,{}^3\!P_2$}
 \newcommand{\Pone}{$2\,{}^3\!S_1 \rightarrow 2\,{}^3\!P_1$}
 \newcommand{\Pzero}{$2\,{}^3\!S_1 \rightarrow 2\,{}^3\!P_0$}

\draft 
\def\orcid#1{\kern .08em\href{https://orcid.org/#1}{\includegraphics[keepaspectratio,width=0.7em]{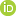}}}

\begin{document}


\title{Collinear laser spectroscopy of highly charged ions produced with an electron beam ion source} 



\author{P.~Imgram\orcid{0000-0002-3559-7092}}
\email{pimgram@ikp.tu-darmstadt.de}
\affiliation{Institut f\"ur Kernphysik, Departement of Physics, Technische Universit\"at Darmstadt, Schlossgartenstraße 9, 64289 Darmstadt, Germany}

\author{K.~König\orcid{0000-0001-9415-3208}}
\affiliation{Institut f\"ur Kernphysik, Departement of Physics, Technische Universit\"at Darmstadt, Schlossgartenstraße 9, 64289 Darmstadt, Germany}
\affiliation{Helmholtz Research Academy 
Hesse for FAIR, Campus Darmstadt,
Schlossgartenstr.\ 9, 64289 Darmstadt}

\author{B.~Maaß\orcid{0000-0002-6844-5706}}
\altaffiliation{current address: Physics Division, Argonne National Laboratory, 9700 S Cass Ave, IL 60439 Lemont, USA}
\affiliation{Institut f\"ur Kernphysik, Departement of Physics, Technische Universit\"at Darmstadt, Schlossgartenstraße 9, 64289 Darmstadt, Germany}

\author{P.~Müller\orcid{0000-0002-4050-1366}}
\affiliation{Institut f\"ur Kernphysik, Departement of Physics, Technische Universit\"at Darmstadt, Schlossgartenstraße 9, 64289 Darmstadt, Germany}

\author{W.~Nörtershäuser\orcid{0000-0001-7432-3687}}
\affiliation{Institut f\"ur Kernphysik, Departement of Physics, Technische Universit\"at Darmstadt, Schlossgartenstraße 9, 64289 Darmstadt, Germany}
\affiliation{Helmholtz Research Academy 
Hesse for FAIR, Campus Darmstadt,
Schlossgartenstr.\ 9, 64289 Darmstadt}


\date{\today}

\begin{abstract}
Collinear laser spectroscopy has been performed on He-like C$^{4+}$ ions extracted from an electron beam ion source (EBIS). In order to determine the transition frequency with the highest-possible accuracy, the lineshape of the fluorescence response function was studied for pulsed and continuous ion extraction modes of the EBIS in order to optimize its symmetry and linewidth. We found that the best signal-to-noise ratio is obtained using the continuous beam mode for ion extraction. Applying frequency-comb-referenced collinear and anticollinear laser spectroscopy, we achieved a measurement accuracy of better than 2\,MHz including statistical and systematic uncertainties. The origin and size of systematic uncertainties, as well as further applications for other isotopes and elements are discussed.   
\end{abstract}


\maketitle 

\section{Introduction}
Precision measurements of fundamental constants and tests of symmetries and interactions have had profound implications for our understanding of nature and the development of theoretical concepts in atomic, nuclear and particle physics. Quantum electrodynamics (QED) plays an important role since it was the prototype of a relativistic quantum field theory and as such a role model for what we now know as the Standard Model. Particularly, few-electron systems that are nowadays available as highly charged ions for precision spectroscopy are very good probes to test QED in strong fields \cite{Klaft1994,Seelig1998}, to study electron correlation effects in simple systems \cite{Winters2011}, and provide highly forbidden transitions that can serve as optical clocks \cite{Kozlov18} as has been very recently demonstrated \cite{King2022}, as probes for variations of fundamental constants, particularly the fine structure constant $\alpha$ \cite{Kozlov18}, or other searches for physics beyond the Standard Model \cite{Safranova2018}. For most of the studies mentioned above, additional effects caused by the finite nuclear size and the nuclear magnetization distribution are detrimental to the goal of the experiments. Therefore, Shabaev et al.\ have suggested to use specific isonuclear differences to eliminate some of the finite nuclear-size effects \cite{Shabaev2001} as it was used, \textit{e.g.}, in \cite{Ullmann2017}, while Paul et al. recently proposed to study transitions between circular Rydberg states in muonic atoms where nuclear contributions are vanishing while bound-state QED effects are still large \cite{Paul2021}.
However, if QED contributions are sufficiently well understood and have been validated in test measurements, the calculations can be used to determine nuclear parameters that are connected to the nuclear distributions, see \textit{e.g.}, \cite{Udem97,Pohl2010,Udem2018} for the size of the proton from transition frequencies or for the nuclear magnetization radius based on experimental hyperfine structure splittings in highly charged ions \cite{Crespo1998,Karpeshin2015}. Even though the effects are small, their influence is sufficient to contribute to several significant digits of achievable measurement accuracy and can be used as a very clean probe of low-energy Quantum Chromodynamics (QCD) properties and nuclear structure \cite{Kozlov18}.

Light nuclei present unique systems, as they can be accurately calculated using \textit{ab initio} methods based on systematic nuclear forces \cite{Epelbaum2009,Hammer2013,Hergert2016,Hebeler2021}. It is also a region in which spectacular nuclear structure effects arises - the so-called halo nuclei \cite{Tanihata1985,Tanihata1996}. The charge radius of the most prominent halo nucleus $^{11}$Li \cite{Sanchez2006} is still limited by the knowledge of the stable $^6$Li reference radius \cite{Nortershauser2011}. This is even worse for boron, for which the uncertainty of the charge radius of the stable isotopes is a serious obstacle for the determination of the charge radius of the proton-halo candidate $^8$B as well as for tests of the latest generation of nuclear structure calculations \cite{Ryberg2014,Maaß2019}. The importance of further progress in this field has also been highlighted by the still not fully solved ``proton-radius puzzle'' \cite{Pohl2010,Karr2020}. With respect to this work, the remaining discrepancies between radii extracted from electronic and muonic systems are of interest, whereas the discrepancies to electron scattering results and among different electron scattering experiments are a topic of their own. For a recent review see, \textit{e.g.}, \cite{Gao2022}. 
Both aspects, QED tests and determination of nuclear structure, are strongly intertwined and measurements on different elements, isotopes, and charge states are required to prove consistency. This also includes exotic systems like muonic atoms for which measurements are planned in ion traps \cite{Schmidt2018} or using a new generation of microcalorimeters \cite{Antognini2022,Antognini2020}. 

A particularly interesting case in this respect is an all-optical determination of the nuclear charge radius $R_\mathrm{c}$ in few-electron systems. The general idea of an all-optical charge radius measurement is the determination of the difference between the theoretical transition frequency $\nu_\mathrm{point}$ of an ionic or atomic system calculated assuming a point-like nucleus and the experimental transition frequency $\nu_\mathrm{exp}$ of the real system. If all other effects are covered by the atomic structure calculations, the difference is only given by the finite nuclear-size effect. If, finally, the sensitivity of the transition to the charge radius $F$ in MHz/fm$^2$ is also known, the charge radius can be determined according to
\begin{equation}
    R_\mathrm{c} = \langle r^2 \rangle^{1/2} = \sqrt{\frac{\nu_\mathrm{exp} - \nu_\mathrm{point}}{F}}.
\end{equation}

\begin{figure*}[htb]%
\centering
\includegraphics[width=0.9\linewidth]{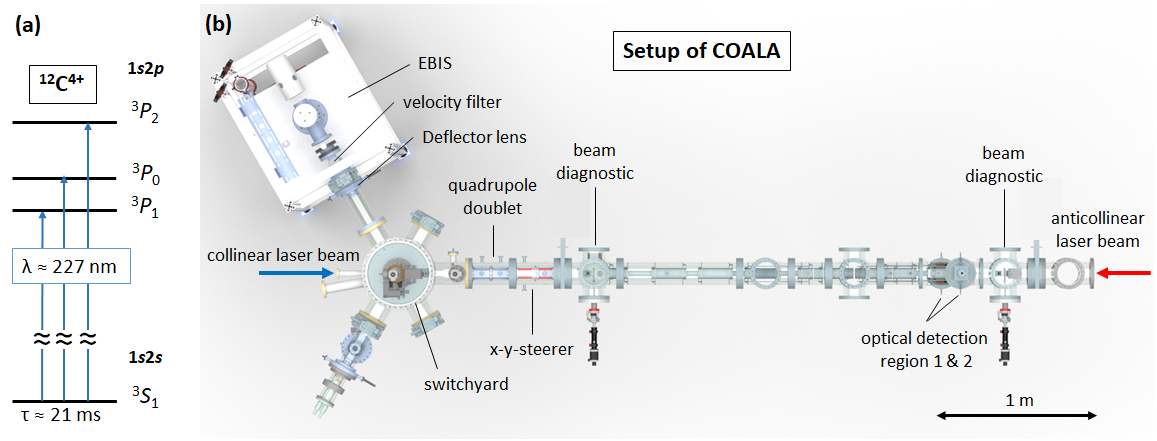}
\caption{Overview of the COALA beamline. Main differences of the nearly 7-m long beamline to the previous version detailed in \cite{König20COALAReview} are the electron beam ion source (EBIS) with an attached velocity filter and the switchyard.}\label{fig:beamline}
\end{figure*}
This approach has been used so far only for hydrogen-like systems, \textit{i.e.}, hydrogen \cite{Udem97, Fleurbaey2018, Grinin2020}, muonic hydro\-gen \cite{Pohl2010}, muonic deuterium \cite{Pohl2016} and muonic helium \cite{Krauth2021, Schuhmann2023}. However, recently a program has been initiated by K.\,Pachucki and V.\,Yerokhin with the goal to determine charge radii using the $^3\!S\, \rightarrow\, ^3\!P$ transitions of (ortho-)helium and helium-like ions \cite{Yerokhin2018}.
The corresponding transition wavelengths are in the optical range up to C$^{4+}$, however, laser spectroscopy experiments are challenging since those start from a metastable state that has to be populated first. Additionally, the lifetime of the $^3\!S$ state decreases rapidly with $Z^{-12}$, reducing the lifetime of the $1s2s\,^3\!S_1$ state from 2.2\,hours in He to 21\,ms in C$^{4+}$ as listed in Tab.\,\ref{tab:lifetime}.

Laser spectroscopy can be performed \textit{in situ} with the production as demonstrated for other applications in Ar$^{13+}$ \cite{Mäckel2011}, Fe$^{13+}$ \cite{Schnorr2013} and I$^{7+}$ \cite{Kimura2023} in an electron beam ion trap (EBIT), however, due to the prevailing high temperatures and correspondingly large Doppler widths, this approach delivers only a precision on the 100 parts-per-billion (ppb) level. Alternatively, the ions can be extracted from the production zone, cooled and transferred into a Penning \cite{Gruber2001,Hobein2011,Andelkovich2013,Egl2019} or a Paul trap \cite{Schmoeger2015, Micke2020, King2022}, improving the precision by several orders of magnitude.
However, this technique is not applicable for He-like systems beyond Be$^{2+}$, since the lifetime degrades to a few 100\,ms and less. Thus, the investigation of He-like B to N isotopes has similar conditions as laser spectroscopy of short-lived radioactive nuclei, for which collinear laser spectroscopy was specifically developed \cite{Kaufman1976,Schinzler78}.

Therefore, we have coupled an electron beam ion source (EBIS) to a collinear laser spectroscopy setup. We produced and extracted C$^{4+}$ ions and performed frequency-comb referenced quasi-simultaneous collinear and anticollinear laser spectroscopy to determine accurate transition frequencies. While those are published in \cite{Imgram23_PRL}, this paper covers a detailed insight into the experimental setup, a comparison of the different EBIS production modes and a discussion of systematic uncertainties.

\begin{table}[hbt]
   \centering
   \caption{Lifetime $\tau$ and transition wavelength $\lambda$ of the metastable  $2\,{}^3\!S_1$ state in He-like ions \cite{NIST_ASD}.}
   \begin{tabular}{ccc}
   \hline
       \hspace{2cm} & \textbf{$\tau (2\,{}^3\!S_1)$} & \textbf{$\lambda(2\,{}^3\!S \rightarrow 2\, ^3\!P)$} \\\hline
           He   &  2.2\,h & 1082\,nm \\
           Li$^+$   &   50\,s & 548\,nm \\
           Be$^{2+}$ & 1.8\,s & 372\,nm \\
           B$^{3+}$ & 150\,ms & 282\,nm \\
           C$^{4+}$ & 21\,ms & 227\,nm \\
           N$^{5+}$ & 3.9\,ms & 190\,nm \\
           \hline
   \end{tabular}
   \label{tab:lifetime}
\end{table}

\section{Experimental setup}
The Collinear Apparatus for Laser Spectroscopy and Applied Science (COALA), situated at TU Darmstadt/Germany, has been extended with an electron beam ion source (EBIS) to allow for collinear laser spectroscopy of highly-charged ions, especially at low masses. An illustration of the experimental setup is shown in Fig.\,\ref{fig:beamline}. COALA has proven to be a valuable setup for high-precision collinear laser spectroscopy. Rest-frame transition frequencies of allowed dipole transitions in Ba$^+$ and Ca$^+$ with an accuracy comparable to ion trap measurements were extracted to investigate a puzzling behavior in the atomic structure of earth-alkaline elements \cite{Imgram19, Müller20}. Furthermore, the apparatus can be used to perform laser-assisted high-voltage measurements \cite{Krämer18}. The main beamline of COALA remained unchanged for the investigations presented here and is described in detail in \cite{König20COALAReview}. In this paper, we concentrate on an extended description of the newly installed elements to generate a beam of highly charged ions  and give only a brief overview of the main beamline, as far as it is required to comprehend the presentation of the experiment and the discussion of the systematic uncertainties arising from beam properties.

\subsection{Dresden EBIS-A}
\begin{figure}[htb]%
\centering
\includegraphics[trim=65mm 65mm 63mm 52mm,clip,width=1.0\linewidth]{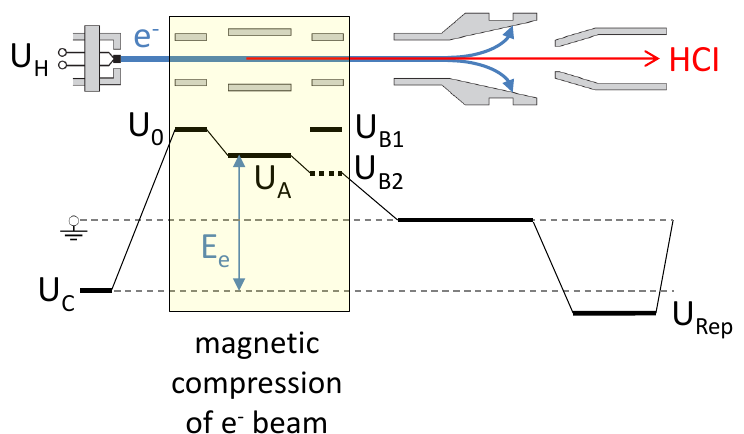}
\caption{Schematic illustration of the potentials inside the EBIS. Taken and modified from \cite{Zschornack14}.}\label{fig:ebis_scheme}
\end{figure}

The Dresden EBIS-A is a room-temperature electron beam ion source from DREEBIT GmbH. It produces highly charged ions through electron impact ionization. The operation principle of an EBIS is extensively explained in, \textit{e.g.}, \cite{Shirkov96, Zschornack14}.
\newline
Here, we will concentrate on those aspects that are relevant for the ion beam properties affecting the resonance lineshape. A schematic illustration of our EBIS and the applied voltages are depicted in Fig.\,\ref{fig:ebis_scheme}. An iridium-cerium cathode generates an electron beam of up to 120\,mA which is subsequently accelerated from the cathode potential $U_\mathrm{C} \approx -2150$\,V into three drift tubes forming the ion trap. Afterwards, the electron beam is repelled by a negative voltage $U_\mathrm{Rep} < U_\mathrm{C}$ and guided onto a water cooled electron collector. Electron beam compression is realized by an axially symmetric magnetic field created and formed by two NbFeB permanent magnet rings and soft iron parts producing an on-axis magnetic field strength of $\approx 620$\,mT.
Atoms inside the electron beam are efficiently ionized through electron impact ionization. Positive ions produced in the central drift tube, which has an effective length of 60\,mm and an inner diameter of 5\,mm, are axially trapped by the potential $\delta U_\mathrm{trap} = U_\mathrm{B1} - U_\mathrm{A}$. The radial ion trap is formed by the negative space charge of the electron beam. The central potential $U_\mathrm{A}$ was usually set to approximately 10.5\,kV which was a compromise of having a high starting potential for the ions while limiting the recurrences of discharges inside the source.

The easiest way of feeding the EBIS with the element of interest is through leakage of a gaseous sample into the drift tube section. For the production of C$^{4+}$, propane gas (C$_3$H$_8$) was used. The base pressure inside our EBIS was $p_\mathrm{base} = 8\cdot10^{-10}$\,mbar and the typical feeding gas pressure $p_\mathrm{gas} = 6\cdot10^{-8}$\,mbar. The molecules are disaggregated by electron impact ionization, the formed ions are trapped inside the electron beam and subject to further collisional processes with electrons, atoms and other ions, increasing or decreasing their charge state. In the most common EBIS operation mode, ions are pulsed out by rapidly lowering the third electrode from the potential $U_\mathrm{B1}$ to $U_\mathrm{B2}$.
The trapping or breeding time $\tau_\mathrm{breed}$ of the ions determines the resulting charge-state distribution of the ensemble. An optimum for the production of C$^{4+}$ was found for $t_\mathrm{breed} = 15$\,ms. 
The ejected ion bunch contained roughly $8\cdot10^7$ C$^{4+}$ ions in 8\,\textmu s.

\begin{figure}[tbh]%
\centering
\includegraphics[width=1.0\linewidth]{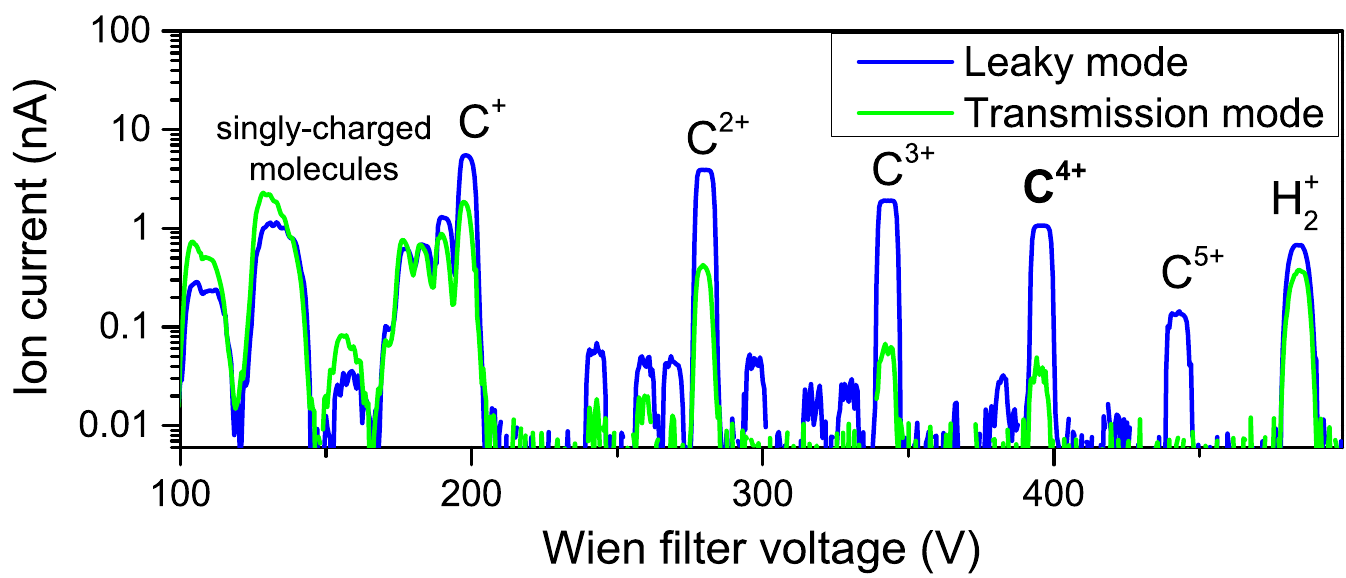}
\caption{Comparison between leaky mode ($U_\mathrm{A} < U_\mathrm{B1} < U_0$) and transmission mode ($U_\mathrm{B1} \leq U_\mathrm{A} < U_0$) production with propane gas (C$_3$H$_8$) through scanning of the Wien filter voltage and recording of the transmitted ion current. Besides the prominent charge states of carbon, also singly-ionized molecules and small amounts of O and N charge states are produced in these modes.}\label{fig:leak_prod}
\end{figure}

Alternatively, a continuous extraction of ions is realized by choosing a lower and static $U_\mathrm{B1}$ with respect to $U_0$ but above $U_\mathrm{A}$ ($U_\mathrm{A} < U_\mathrm{B1} < U_0$) so that ions can overcome the rear-wall potential once the space charge of the ions inside the trap is sufficiently large. Figure\,\ref{fig:leak_prod} shows a comparison between the charge state production in an open trap ($U_\mathrm{B1} \leq U_\mathrm{A} < U_0$, transmission mode) and the leaky mode. The spectrum is generated by a Wien-filter scan. While the transmission mode produces mainly singly-ionized molecules and low charge states, the leaky mode produces significant amounts of multiply charged ions. Of particular importance is the peak of C$^{5+}$, since the metastable state in C$^{4+}$ is most efficiently populated by electron capture and charge exchange reactions of these ions. In leaky mode, a continuous C$^{4+}$ ion beam with a particle current of typically 350\,ppA (particle pico ampere) was achieved. We finally note that small amounts of O and N charge states are generated from residual gas.
A dedicated investigation of spectroscopic differences between pulsed and continuous modes is described in Sec.\,\ref{sec:spectra}.

After release, the ions are accelerated from the start potential $U_\mathrm{start} = U_\mathrm{A}$ towards ground potential and pass a Wien filter integrated in the EBIS. The filter consists of a permanent magnet ($B = 500$\,mT) and a variable electric field. By choosing a matching electric field, a specific mass-to-charge ratio can be selected. Ions of other $m/q$-ratios are deflected horizontally and blocked by a 2-mm aperture. This helps to optimize the production of a specific charge state and reduces contaminants in the ion beam that can lead to an increase of unwanted collisions or to space-charge effects.
\subsection{Switchyard}
\begin{figure}[t]%
\centering
\includegraphics[width=0.9\linewidth]{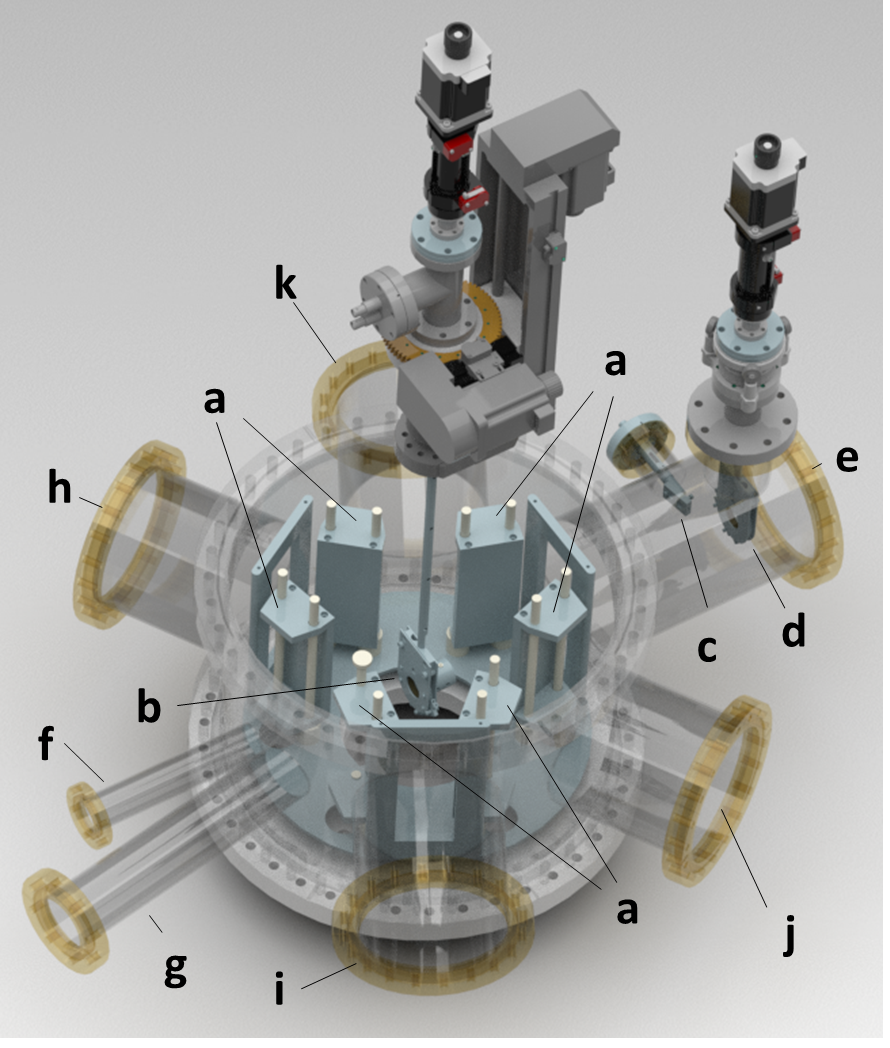}
\caption{The new switchyard for COALA adapted from a design used at the Extra Low ENergy Antiproton (ELENA) ring, CERN \cite{ELENADesignReport}. Details can be found in the text. (a) deflectors, (b) rotatable and movable Faraday cup with an iris, (c) Faraday cup for entry port (g),  (d) iris defining the entrance into the main beamline, (e)  connection point to the collinear beamline, (f) laser-entry port, (g) 10\degree-entry port, (h,i,j,k) main connection ports. }\label{fig:switchyard}
\end{figure}

The production of highly charged ions with an EBIS requires an ultra-high vacuum below $10^{-9}$\,mbar inside the source. This makes it mandatory to bake out the whole source over several days after venting. Together with the heavy frame including the pumping stages, this source is not readily interchangeable like the other ion sources that have been used at COALA so far. Since the setup described in \cite{König20COALAReview} had only one ion-source port, a new switchyard for multiple sources was designed to ensure rapid switching of ions from different sources.
\newline
The design is based on a switchyard from the Extra Low ENergy Antiproton (ELENA) ring at CERN for which details can be found in \cite{ELENADesignReport}. Figure\,\ref{fig:switchyard} shows the adapted COALA version without the cover flange for better visibility. The main difference to the ELENA version is the equally spaced 120\degree-arrangement of the bending electrodes (a) while keeping all other symmetries from the original design. This offers additional flexibility since ions from the EBIS at port (h) or another ion source at port (i) can also be used in possible new experiments to be mounted at (j) or (k) instead of being delivered into the main beamline (e). Furthermore, ion sources such as a liquid metal ion source (LMIS) or a Penning ion source (PIG) at port (j) or (i) can be used to feed the EBIS with ions that cannot be produced from vaporized compounds, for example beryllium. 
Additionally, the 10\degree-port (g) from the original COALA setup was retained. Ions delivered from this port can only be directed into the main beamline (e) with the two steerer electrodes (a) into exit port (e).

In order to measure the ion current from each ion source and to estimate the ion beam size, a Faraday cup (b) with an iris diaphragm in front is installed in the center of the chamber. This cup can be rotated towards each port and moved in and out with a linear $z$-stage. Since the central axis of port (g) does not pass the center of the chamber, another Faraday cup (c)  for the 10\degree-port has been implemented in a straight line. Another iris diaphragm (d) which is aligned to the intended beam axis can be used to ensure a central entry into the quadrupole-doublet of the main beamline. Opposite to the main beamline port (e) is the laser entry port (f).

\subsection{Main beamline}
The main beamline (see Fig.\,\ref{fig:beamline}) starts with an electrostatic quadrupole-doublet to compensate the double-focusing behavior of the switchyard and to recollimate the ion beam. A subsequent \textit{x}-\textit{y}-steerer is used to position the ion beam. The necessary axis for the collinear or anticollinear superposition of the laser and ion beam is defined by two iris diaphragms inside the two beam diagnostic stations. Furthermore, a Faraday cup and a multi-channel plate (MCP) phosphorus-stack is available in each station to observe and monitor the ion beam. The actual laser spectroscopy is performed in the optical detection region (ODR). 

When the C$^{4+}$ ions are excited from the metastable $2\,{}^3\!S_1$ to the $2\,{}^3\!P_J$ states which have a lifetime of 17\,ns, fluorescence light is emitted at the excitation wavelength when decaying back into the $^3\!S_1$ state. The fluorescence photons are collected by two elliptical mirrors (ODR1 \& ODR2) and detected by photomultiplier tubes \cite{MaaßArxiv}. Each photon count is registered with 10-ns resolution by the FPGA-based data acquisition (DAQ) system \cite{KaufmannDiss}.

\subsection{Laser system}
The laser system to produce the necessary 227-nm light consists of a Ti:sapphire (Ti:Sa) laser (Sirah Matisse 2 TS) pumped by a 20-W frequency-doubled Nd:YAG laser (Spectra-Physics Millenia eV) and two following frequency-doubling units (Sirah WaveTrain 2). Two identical systems are available for both directions to perform collinear and anticollinear measurements in fast iterations. Laser collimation and a Gaussian beam profile are realized with a spatial filter behind the second frequency doubling stage. Then, the UV light is transported from the laser laboratory to the beamline through air. With a second telescope in front of the far end of the beamline a good spatial overlap of the two beam profiles with a beam diameter of about 1\,mm inside the ODR is ensured.

The laser frequency of the Matisse is stabilized to a tunable reference cavity through the side-of-fringe stabilization technique. This results in a short-term spectral linewidth of the fundamental frequency of roughly 200\,kHz. To provide long-term stability over a measurement cycle (5 - 10\,min), the fundamental laser frequency is measured and stabilized by a Menlo-Systems FC1500-250-WG frequency comb, whose frequencies are generated with respect to a GPS-disciplined quartz oscillator. The typical Allan deviation is approximately 20\,kHz in time spans of a few minutes and even less for longer time spans.

The laser light is linearly polarized throughout the beam path. To ensure the linear polarization in the laser spectroscopy process, Rochon prisms with a distinction ratio of 10000:1 were placed in front of the beamline for both directions.

\section{Resonance spectra}\label{sec:spectra}
In collinear laser spectroscopy, the resonance condition of an atomic transition is shifted for a counter-propagating (anticollinear, a) and co-propagating (collinear, c) laser beam by the relativistic Doppler effect due to the velocity $\beta = \upsilon/c$ of the ions according to $ \nu_\mathrm{c/a} = \nu_0 \gamma (1 \pm \beta)$ with the Lorentz factor $\gamma = 1/\sqrt{1 - \beta^2}$. This condition can either be met by tuning the laser frequency or by changing the ion velocity through applying typically a few 10\,V to the ODR, which can be floated relative to the rest of the beamline. Usually the latter is easier and faster and therefore the preferred method. The precise determination of the rest-frame transition frequency $\nu_0$ would require precise knowledge of the ion velocity $\beta$ if only $\nu_a$ or $\nu_c$ is measured. However, if the laboratory-frame transition frequencies $\nu_\mathrm{c}$ and $\nu_\mathrm{a}$ are measured in fast iteration, this allows us to directly access $\nu_0^2$ through the geometric mean
\begin{equation}
    \nu_\mathrm{c}\cdot \nu_\mathrm{a} = \nu_0^2 \, \gamma^2 \, (1+\beta) (1-\beta) = \nu_0^2.
    \label{eq:colacol}
\end{equation}
The precise determination of a resonance line center $\nu_\mathrm{c/a}$ depends strongly on the signal-to-noise ratio (SNR), the symmetry of the lineshape and its width. Thus, EBIS production parameters were optimized to achieve a significant population of the $2\,{}^3\!S_1$ state as well as a symmetric and narrow line profile. During these optimizations, a single anticollinear 1-mW laser beam was used and the strongest fine-structure transition \Ptwo ~of $^{12}$C$^{4+}$ was studied. First, the more commonly used bunched mode of the EBIS was studied and then compared to continuous-beam operation of the EBIS.

\subsection{Bunched beam}
\begin{figure}[b]%
\centering
\includegraphics[width=1\linewidth]{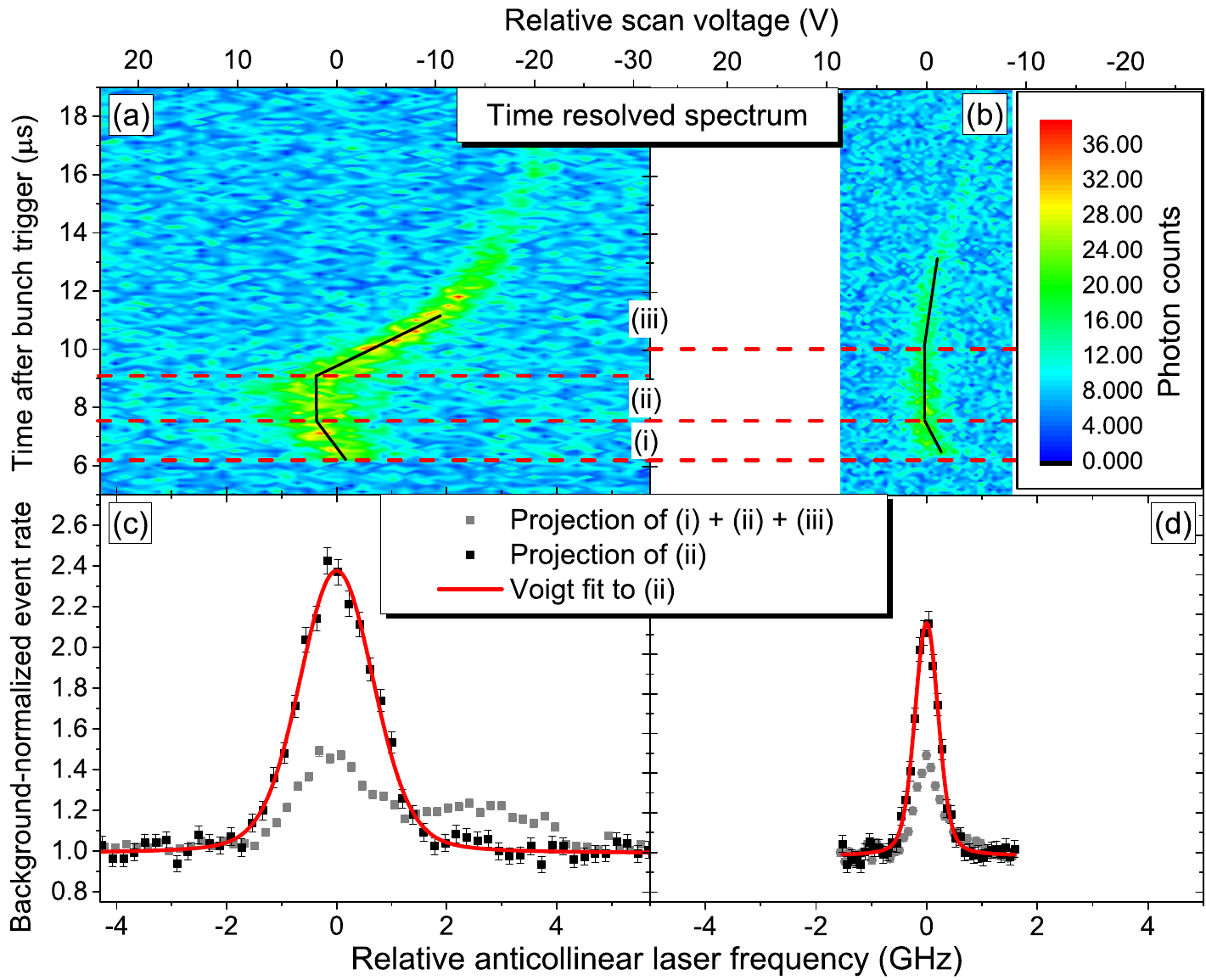}
\caption{Time-resolved spectra for two different EBIS production parameter sets (a) and (b) measured in anticollinear geometry. The projection of the photon events within time window (ii) onto the laser frequency axis is shown in the bottom panels as black squares along with a Voigt fit (red line). The full projection of time window (i) + (ii) + (iii) is depicted as grey squares and exhibits a strong asymmetry for the parameter set (a). The production parameters and the FWHM of the fit are detailed in Tab.\,\ref{tab:bb_production_params}. Further explanation can be found in the text.}\label{fig:bb_spectra}
\end{figure}
In bunched mode, the ions are stored for a certain breeding time $t_\mathrm{breed}$ in the EBIS and afterwards ejected by fast ramping the extraction-electrode potential below the central trap potential. Further important parameters which influence the ion production are the electron current $I_\mathrm{e}$, the trap potential $\delta U_\mathrm{trap}$ and the propane pressure $p_\mathrm{gas}$ inside the EBIS. 
The interplay of all of these parameters defines the electron space charge, the capacity of the trap and the temperature of the ion cloud. Especially the latter is very important since a colder ion cloud results in a smaller linewidth which improves the statistical uncertainty in the determination of the laser-spectroscopic line center. The comparison of many parameter combinations has shown that the amount of produced ions per extraction and the temperature of the ions cannot be optimized simultaneously, but a trade-off has to be made for many parameters. 
Panel (a) and (b) of Fig.\,\ref{fig:bb_spectra} show time-resolved spectra for two different sets of EBIS parameters given in Tab.\,\ref{tab:bb_production_params}. In a time-resolved spectrum, the number of photon counts is depicted color-coded as a function of the scan voltage (top $x$-axis) or the corresponding scan frequency (bottom $x$-axis) and the time after the ejection of the ions from the EBIS ($y$-axis). 
Even though the vertical axis represents the time-of-flight of the ions, it should be noted that the vertical position cannot be directly related to the kinetic energy of the ions since it also depends on the complex extraction process and the position of the ions in the trap while switching the voltage $U_\mathrm{B1}$. The velocity of the ions is rather encoded in the position of the resonance along the $x$-axis because this represents the Doppler-shift. 

The black line emphasizes the change of the resonance frequency with time. In this view, the energy distribution of the ion bunch is visualized in a time-resolved way from bottom to top, where ions that are at resonance at higher frequencies have less kinetic energy than ions in resonance at lower frequencies. Therefore, one can analyse the ion ejection behavior and identify roughly three different parts (i), (ii), and (iii) separated by the red dashed lines.
In part (i) and (iii), the ion energy changes with time whereas it is roughly constant in part (ii). For laser spectroscopy, a time-independent behavior is required to allow for a precise and accurate determination of the resonance frequency. It is obvious that a projection of all fluorescence events onto the frequency axis leads to a strongly asymmetric resonance profile, as depicted in gray in the bottom panels of Fig.\,\ref{fig:bb_spectra}. Restricting the projection to the time period (ii), in which the resonance frequency is constant, provides a symmetric and narrower resonance signal as shown by the black data points. Please note that the background-normalized count rate is depicted in Fig.\,\ref{fig:bb_spectra} (c) and (d). The smaller peak intensity for the full projection (i + ii +  iii) requires a longer acceptance window along the time axis and, thus, collects more background. 

 \begin{table*}[htb]
    \centering
    \caption{EBIS production parameters, signal-to-noise ratio (SNR), full width at half-maximum (FWHM) of the resonance lineshape and statistical uncertainty of the line center $\Delta \nu_\mathrm{center}$ of the spectra in Figs.\,\ref{fig:bb_spectra} and \ref{fig:bb_vs_cb}. The measurement time for each spectrum was roughly 5\,min.}
    \begin{tabular}{c|c|c|c|c||c|c|c}
         & $I_\mathrm{e}$ (mA) & $\delta U_\mathrm{trap}$ (V) & $p_\mathrm{gas}$ (mbar) & $t_\mathrm{breed}$ (ms) & SNR & FWHM (GHz) & $\Delta \nu_\mathrm{center}$ (MHz) \\\hline 
        (A) & 80 & 71 & $6\cdot10^{-8}$ & 15 & 58 & 1.59(4) & 16.6\\
        (B) & 25 & 26 & $6\cdot10^{-8}$ & 15 & 51 & 0.50(1) & 5.7\\
        (C) & 25 & 26 & $2\cdot10^{-8}$ & 15 & 17 & 0.97(4) & 17.8\\
        (D) & 80 & 171 & $6\cdot10^{-8}$ & leaky & 59 & 0.168(2) & 0.7\\
    \end{tabular}
    \label{tab:bb_production_params}
\end{table*}
We suppose that the time-dependence in part (i) is caused by ions that can already leave the trap while the potential of the extraction electrode is still changing. Hereby, the electrode acts unintentionally as an elevator drift-tube where the ions lose energy compared to the ``main bunch'' in part (ii) and the resonance is therefore shifted to higher frequencies. The strongly tilted tail in part (iii), now drifting towards lower ion energies is ascribed to a changing space-charge potential inside the electron beam. In an empty trap, it lowers the nominal start potential $U_\mathrm{start}$. 
At the time of ejection, however, the positive charge of the ion cloud partly compensates the negative electron potential. Therefore, the first ions start on a potential closer to $U_\mathrm{A}$. After some time, when the main part of the ion cloud has left the trap, the negative electron space-charge potential is less compensated and later ions therefore start on a reduced potential. 
This leads to a reduced kinetic energy of these ``tail ions'' after the acceleration against the ground potential. This explanation is supported by the result shown in Fig.\,\ref{fig:bb_spectra}(b), where the electron current $I_\mathrm{e}$ was reduced from 80\,mA to 25\,mA and with it the space-charge potential of the electrons. Consequently, the space-charge induced tail is less prominent and the frequency shift is strongly reduced.

It is obvious that less ions are produced in total with the lower electron current, while the temperature of the stored ion cloud is reduced due to less collisional heating. The latter is directly reflected in the spectral linewidth since the temperature-induced longitudinal energy distribution of the ion cloud is the main reason for the width of the resonance spectrum. For a better comparison, a Voigt profile was fitted to the projections of the time window (ii) in Fig.\,\ref{fig:bb_spectra} and the corresponding signal-to-noise ratio (SNR), full width at half maximum (FWHM) and statistical line-center uncertainty $\Delta \nu_\mathrm{center}$ are listed in Tab.\,\ref{tab:bb_production_params} for different production conditions. The strong reduction of the FWHM from 1.6\,GHz to 0.5\,GHz improves the statistical line center uncertainty from 16.6\,MHz to 5.7\,MHz. Even though the reduced electron current leads to a proportionally smaller ion yield, the SNR is similar for both parameter sets since those ions are compressed into a smaller frequency range. The smaller linewidth thus compensates a part of the signal loss at the peak center. This demonstrates the advantage of a cooled ion bunch.

The electron current is not the only parameter which influences the ion temperature. When the axial trap depth $\delta U_\mathrm{trap}$ is lowered, hot ions start to leave the trap and energy is removed from the thermal equilibrium which cools the remaining ions. This so-called evaporative cooling has been employed in panel (b) of Fig.\,\ref{fig:bb_spectra} to additionally reduce the ion temperature. However, if $\delta U_\mathrm{trap}$ is too shallow, the resonance signal is reduced due to the ion loss. Therefore, the trap depth together with the ion current are trade-off parameters and have been tuned carefully. The best compromise for pulsed extraction was found in the parameter set (B) of Tab.\,\ref{tab:bb_production_params} shown in Fig.\,\ref{fig:bb_spectra}(b).

In contrast, a high propane gas pressure $p_\mathrm{gas}$ improves the production of C$^{4+}$ and cools the ion cloud at the same time. Therefore, it was always set to the maximal value of $p_\mathrm{gas} = 6 \cdot 10^{-8}$\,mbar given by the technical limit of stable operation with rare high-voltage sparks. The breeding time was usually kept at $t_\mathrm{breed} = 15$\,ms since its impact on the C$^{4+}$ production is much larger than on the linewidth.

The experimental determination of the ratio between metastable ions and ground-state ions is difficult as many contributing factors such as the ion-beam -- laser-beam overlap and the absolute photon detection efficiency are elusive. An alternative is the calculation and comparison of the different process rates contributing to the production of C$^{4+}$ in the EBIS \cite{Shirkov96, Zschornack14}. 
The dominant effect in the production of metastable C$^{4+}$ ions is charge exchange $\mathrm{C}^{5+} + \mathrm{X} \rightarrow \mathrm{C}^{4+}(2\,^3\!S_1) + \mathrm{X}^+$ since other possible population modes such as radiative recombination $\mathrm{C}^{5+} + \mathrm{e}^- \rightarrow \mathrm{C}^{4+}(2\,^3\!S_1) + \gamma$ and electron impact excitation $\mathrm{C}^{4+}(1 \,^1\!S_0) + \mathrm{e}^- \rightarrow \mathrm{C}^{4+}(2\,^3\!S_1) + \mathrm{e}^-$ have much smaller cross-sections for an electron energy around 12\,keV \cite{Tully1978}. We have performed simulations of the production mechanism in the EBIS using the Python \textit{ebisim} package \cite{Pahl_ebisim} to obtain the ratio between the charge exchange rate from C$^{5+}$ to C$^{4+}$ and the total production rate of C$^{4+}$ ions. After 15\,ms -- the experimentally optimized breeding time -- this ratio is roughly 14\%. Taking the multiplicity of the atomic states into account, we expect a fraction of approximately 10\% of the C$^{4+}$ ions in the bunched ion beam to be in the laser-accessible metastable $2\,^3\!S_1$ level.
\subsection{Continuous beam}
\begin{figure}[b]%
\centering
\includegraphics[width=1.0\linewidth]{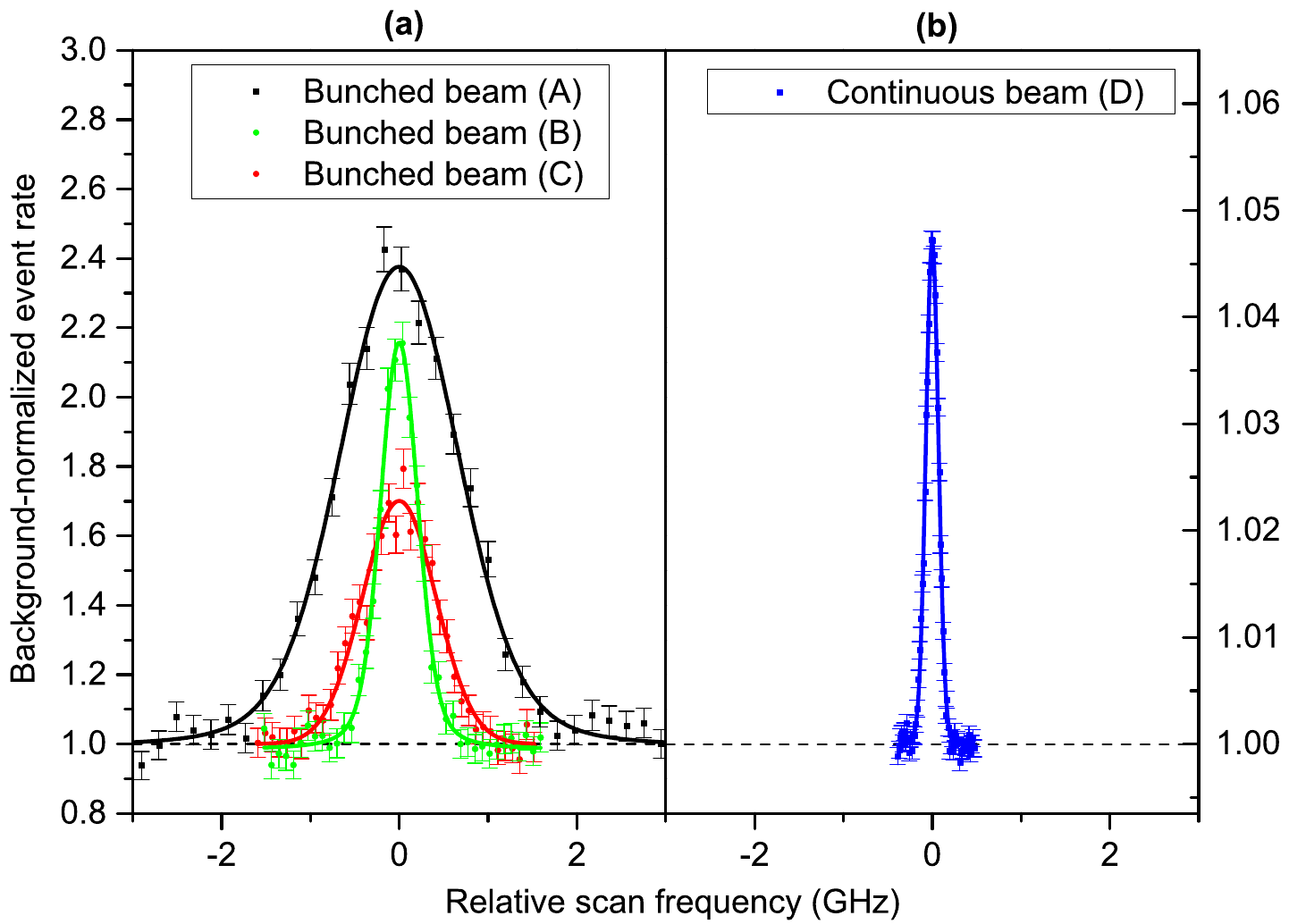}
\caption{Resonance spectra for bunched ion beams (a) with different production settings (A--C) and continuous ion beams (b). Results of the respective Voigt fits are listed in Tab.\,\ref{tab:bb_production_params}.}\label{fig:bb_vs_cb}
\end{figure}

The observation of the $^{12}$C$^{4+}$ resonance in pulsed mode facilitated the signal search in continuous-beam mode. Initially, we were not sure whether the signal intensity would be sufficient in continuous-beam operation since the beam intensity is considerably reduced compared to the peak intensity in the short pulse. Additionally, background reduction by gating the photon counting to the passing of the ion pulse as done in Fig.\,\ref{fig:bb_spectra} does not work anymore. However, a collinear laser spectroscopy resonance was finally observed as shown in the right panel of Fig.\,\ref{fig:bb_vs_cb}, where it is compared to resonance spectra of the bunched-beam mode depicted on the left. For better comparison, both signals are normalized to the background, being set to 1. Hence, the $y$-axis directly reflects the signal-to background ratio. The experimental conditions and spectrum parameters obtained in the fit of the lineshape are summarized in Tab.\,\ref{tab:bb_production_params}. We note that a similar SNR was reached in continuous mode (D) and in pulsed operation optimized for signal intensity (A) within the same measurement time although the signal strength in continuous mode is just 5\% of the background, whereas it is 140\% of the background in bunched mode. The reason for this is twofold: First, the reduced linewidth increases the SNR of the continuous beam spectrum linearly since the signal is concentrated in a smaller spectral range and, secondly, the population of the $2\,^3\!S_1$ state is roughly 4--5 times higher in the continuous beam mode. The latter is explained by more frequent charge-exchange collisions due to the longer trapping time of an individual ion leading from the charge state C$^{5+}$ back to C$^{4+}$. 
The linewidth in the continuous mode is thus reduced from 1.6\,GHz in operation mode (A) to only 0.168\,GHz in continuous mode (D). The improvement of the statistical uncertainty of the line center obtained in a single fit for setting (D) is also striking since it is reduced by more than a factor of 20 from 16.6\,MHz to 0.7\,MHz. Even under the best pulse-mode conditions with respect to the linewidth (B), the linewidth and statistical uncertainty in continuous-beam mode are still improved  by a factor of 3 and 8, respectively. An explanation for the reduced linewidth can be found in the ejection behavior of the ions in the continuous-beam mode. Firstly, no fast-switching potentials of the ejection electrode can smear out the starting potential of the ions like in the bunched-beam mode. Secondly, the space-charge potential of the ion cloud is in a state of equilibrium in the continuous-beam mode which results in a well defined starting potential. Finally, only ions with enough kinetic energy to overcome the static barrier potential will leave the trap. This means that only ions representing a fraction of the full energy distribution in the trap are ejected into the beamline. All effects together result in a reduced energy spread of the ion beam and a narrower observed linewidth.

Although in both modes the natural linewidth of roughly 9\,MHz cannot be resolved, the continuous beam delivers far better conditions for high-precision collinear laser spectroscopy and was therefore the preferred mode for all further measurements and investigations.

\section{Results of the frequency determination}
 \begin{figure}[b]
	\centering
	\includegraphics[width=\linewidth]{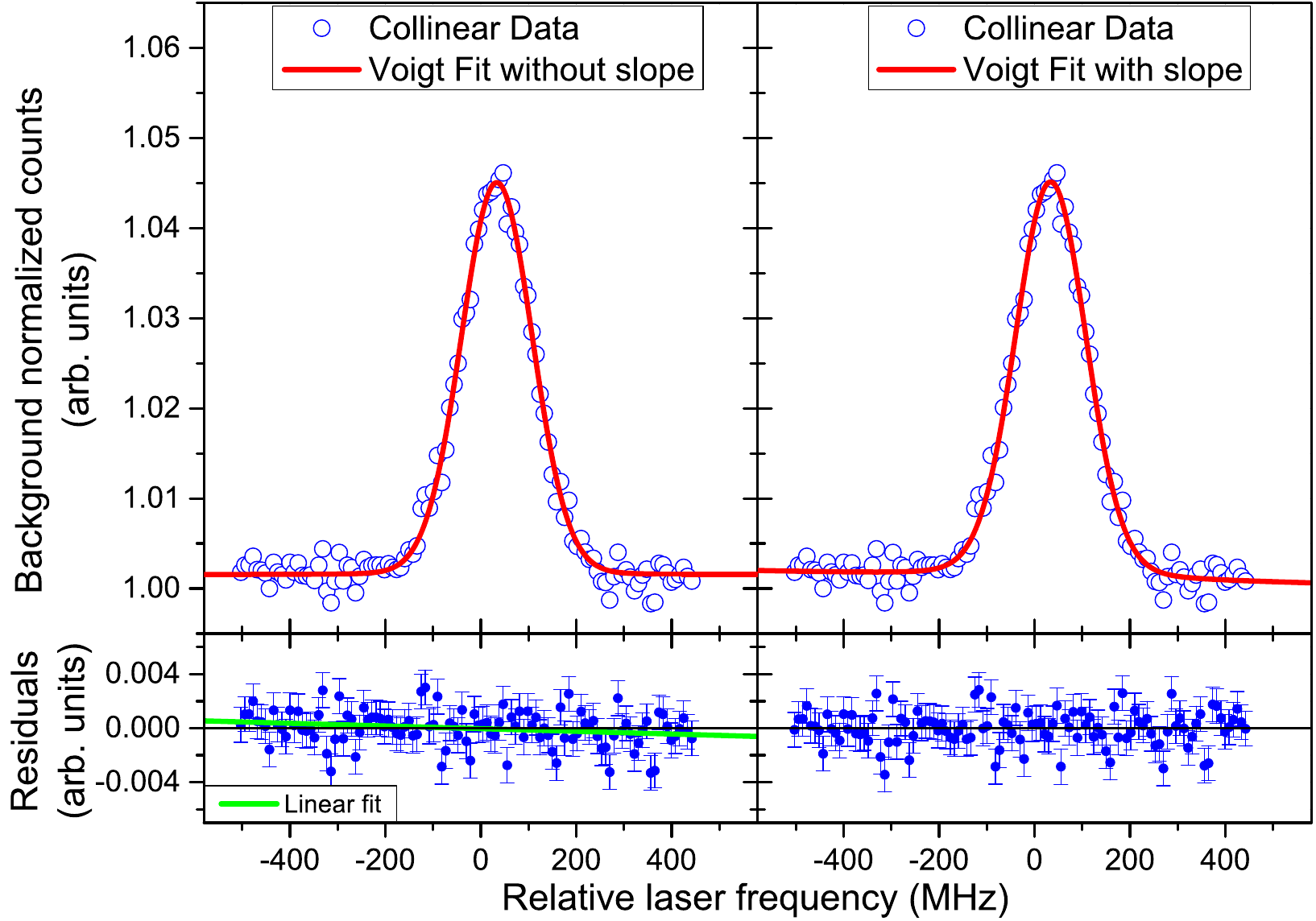}
	\caption{Comparison between a Voigt profile with static background (a) and with a linear background slope (b). The latter removes the slight tilt (green line) in the residuals of the static background fit. Although this shifts the fitted line-center by roughly 0.5\,MHz, the influence on the determination of the rest-frame frequency is almost eliminated if the same fit model is used for both directions.}
	\label{fig:lineShapeComp}
\end{figure}
After a comparison of the EBIS production modes and their spectral line profiles, frequency-comb-referenced quasi-simultaneous collinear and anticollinear laser spectroscopy \cite{Krieger2017, Imgram19, Müller20} has been used to measure the \PJ ~rest-frame transition frequencies $\nu_0$ in $^{12}$C$^{4+}$. We note that we address all possible Zeeman transitions accessible with linearly polarized light simultaneously since the Zeeman splitting induced by the earth magnetic field is considerably less than the natural linewidth of the transition.

In order to extract the central resonance frequencies $\nu_\mathrm{c/a}$ for the collinear ($\nu_\mathrm{c}$) and anticollinear ($\nu_\mathrm{a}$) direction, a function which describes the spectrum must be fitted to the data points. Typically, a Voigt profile, which is the convolution of a Gaussian and Lorentzian profile, is used. In this work, we tested different profiles such as pure Gaussian, a Voigt and a Voigt with added linearly-tilted background. The latter delivered the smallest reduced $\chi^2$ values and the smallest fit uncertainties. Besides the small correction with the slightly tilted background, the resonance lineshape is symmetric and does not show any residual structure as can be seen in the lower trace in Fig.\,\ref{fig:lineShapeComp}(b). The linewidth of the fitted Voigt profile is dominated by the Gaussian contribution due to the energy spread of the ions. A typical Voigt fit yielded a Lorentzian FWHM of $\sim$14\,MHz and a Gaussian FWHM of $\sim$165\,MHz. Therefore, the influence of the laser linewidth and effects such as power broadening can be neglected. The slightly tilted background in some spectra can be explained by variations in the laser power which did not fully average out during the scan. The choice of the fit function obviously influences the value for the extracted center frequencies $\nu_\mathrm{c/a}$. However, performing a full analysis with the different line profiles showed no influence to the averaged rest-frame transition frequency $\overline{\nu}_0$ as long as the same profile is chosen for all measurements.

\begin{figure}[t]
	\centering
\includegraphics[width=1.0\linewidth]{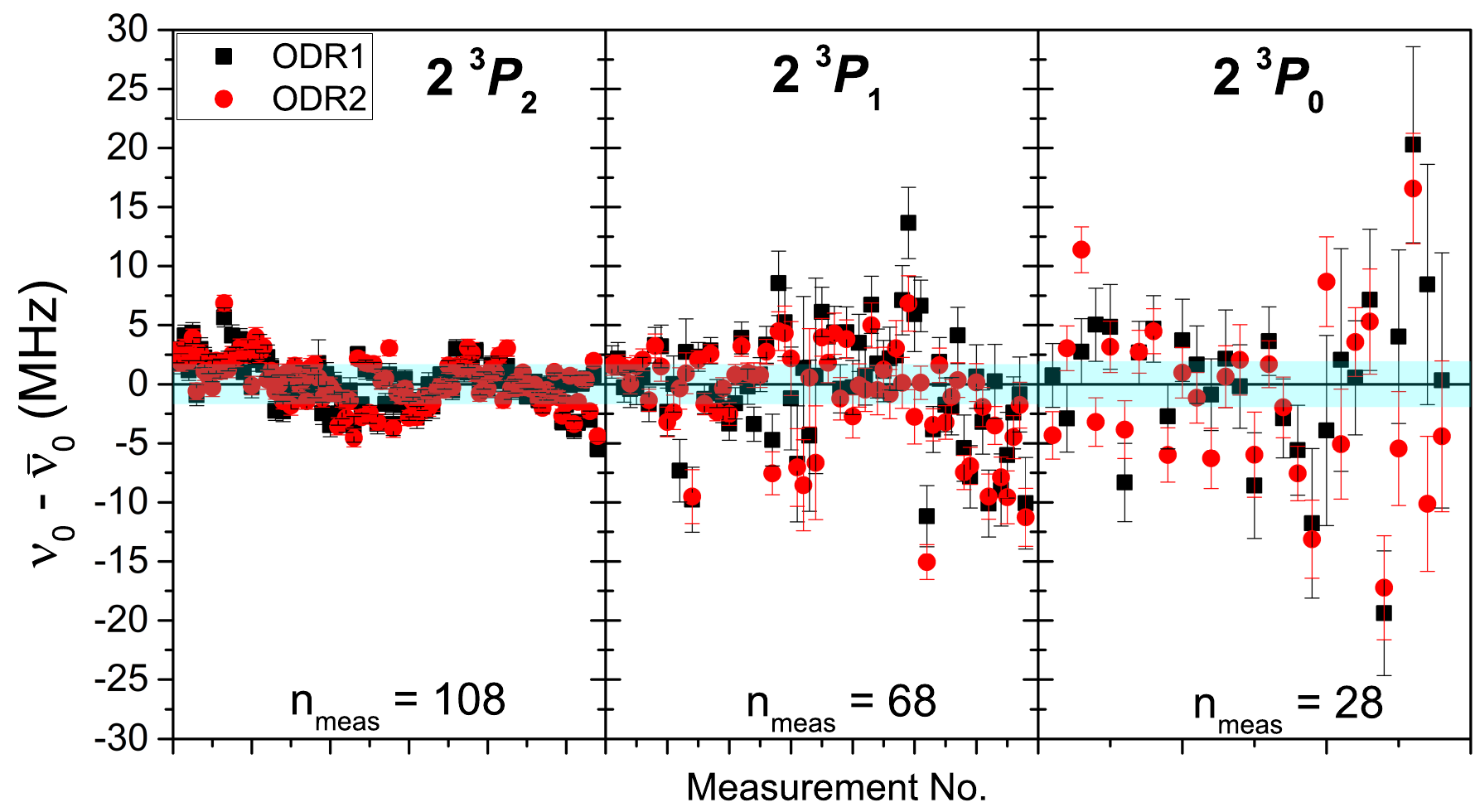}
\caption{Single frequency measurements of the \PJ ~transitions relative to the weighted mean $\overline{\nu}_0$ with their statistical uncertainty. All spectra have been evaluated separately for ODR1 (black squares) and ODR2 (red dots). The shaded blue area indicates the combined systematic (see Sec.\,\ref{sec:uncert}) and statistical uncertainty (standard error of the mean). Note that the precision in all transitions is limited by the systematic uncertainty and additional measurements would not reduce the overall uncertainty.}
	\label{fig:freq_result}
\end{figure}

Each measurement of the rest-frame frequency $\nu_0$ consists of one collinear and one anticollinear measurement with an assigned statistical uncertainty obtained from the combination of the laser frequency uncertainty and the fit uncertainty. 
In Fig.\,\ref{fig:freq_result}, the results of all measurements from the campaign are shown. The data was taken over several days with a separate alignment procedure each day. The values are presented relative to their weighted average $\overline{\nu}_0$. In total, 108 measurements for the \Ptwo, 68 measurements for the \Pone, and 28 measurements for the \Pzero\ transition were carried out. Each measurement was evaluated separately for the two optical detection regions (ODR1 \& 2). 
The blue shaded area marks the combined systematic (see Sec.\,\ref{sec:uncert}) and statistical uncertainty (standard error of the mean) of the weighted average which is also the final value for the respective \PJ ~transition. The precision in all transitions is mainly limited by the systematic uncertainty as discussed below.
We also completed some collinear and anticollinear runs with the bunched ion beam. The mean value of roughly 20 measurements was in good agreement with the continuous-beam measurements. However, the statistical and systematical uncertainties were more than an order of magnitude larger than for the continuous beam and we therefore restricted our investigations to the continuous beam.

\begin{table*}[htb]
    \centering
    \caption{Literature frequency values relative to the rest-frame frequencies $\nu_\mathrm{lit} - \nu_\mathrm{this~work} = \delta\nu_\mathrm{rel} (\,^3$P$_J)$ in the \PJ{} transitions of $^{12}$C$^{4+}$ obtained within this work. The values in brackets denote the uncertainty of the respective literature value. All values are in GHz. The full transition frequencies are provided in \cite{Imgram23_PRL}.}
    \begin{minipage}{0.6\linewidth}
    \begin{ruledtabular}
        \begin{tabular}{l d  d  d  l}
	    & \multicolumn{1}{c}{$\delta \nu_\mathrm{rel} (\,^3$P$_2)$} & \multicolumn{1}{c}{$\delta \nu_\mathrm{rel} (\,^3$P$_1)$} & \multicolumn{1}{c}{$\delta \nu_\mathrm{rel} (\,^3$P$_0)$} & \multicolumn{1}{c}{Ref.}\\
	    \hline
        (E) &-1.3(29.0)    & -7.4(28.9) & 4.5(28.9) & [\onlinecite{Edlen1970}] (Exp.)\\
        (F) &-4.0(4.8) &  & 5.5(4.8) & [\onlinecite{Drake88}] (Theory) \\
        (G) &4.4(3.6) & 2.5(4.2) & 4.0(4.5) & [\onlinecite{Ozawa01}] (Exp.)\\
        (H) &0.03(1.04) & 0.31(1.04) & 0.30(1.04) & [\onlinecite{Yerokhin2010}] (Theory)\\
        (I) &-0.02(13) & 0.41(75) & -0.19(0.27) & [\onlinecite{Yerokhin2022}] (Theory)\\
        \end{tabular}
        \end{ruledtabular}
    \end{minipage}
    \label{tab:freq_comp}
\end{table*}

\begin{figure}[htb]
	\centering
		\includegraphics[width=1.0\linewidth]{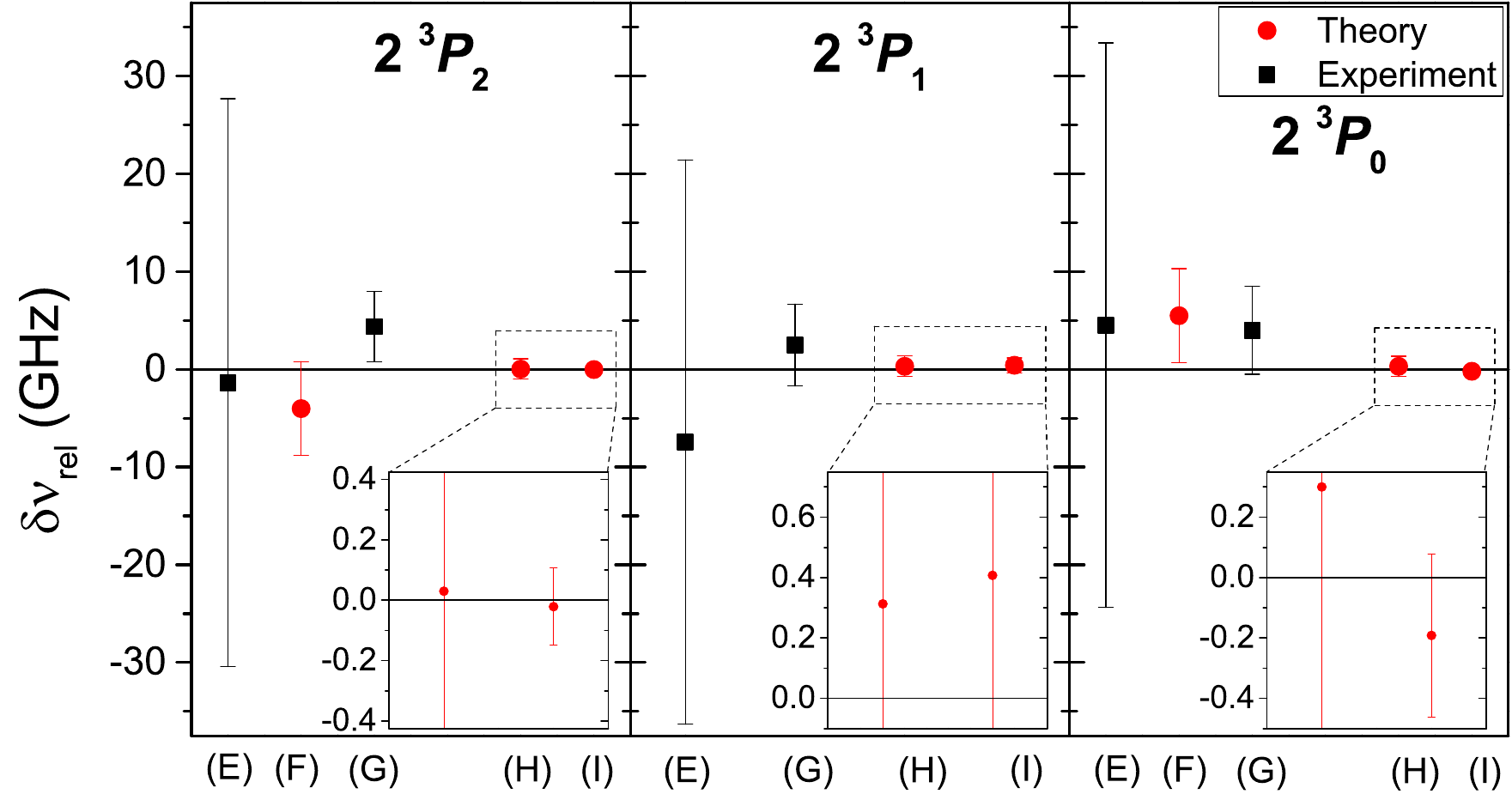}
	\caption{Illustration of the results listed in Tab.\,\ref{tab:freq_comp}. The experimental accuracy has been improved by more than three orders of magnitude compared to the so-far most precise measurement by Ozawa \textit{et al.} (G) \cite{Ozawa01}. Also theory improved by an order of magnitude in a recent publication (I) \cite{Yerokhin2022} compared to (H) \cite{Yerokhin2010}, but it is still two orders of magnitude worse than the experimental accuracy. The uncertainty of this work is thus not visible on this scale.}
	\label{fig:freq_result_comp}
\end{figure}

The differences between transition frequency values from theory and experiment in literature to our results $\delta \nu_\mathrm{rel} = \nu_\mathrm{lit} - \nu_\mathrm{this~work}$ are provided in Tab.\,\ref{tab:freq_comp} and illustrated in Fig.\,\ref{fig:freq_result_comp}. The values in parenthesis denote the uncertainty of the respective literature value since these are in all cases significantly larger than our combined experimental uncertainty of less than 2\,MHz \cite{Imgram23_PRL}. The comparison shows that the most recent non-relativistic QED calculations \cite{Yerokhin2022} refined their accuracy by about one order of magnitude and they agree well within their stated uncertainties with our results, which have uncertainties that are more than 1000 times smaller than the best previous experimental values. The calculated \Pone\ transition frequency has the largest uncertainty due to fine-structure mixing of the $2p\,{}^1$P$_1$ with the $2p\,{}^3\!P_1$ state that have the same angular momentum and parity. Therefore, they have to be treated as quasidegenerate levels in second-order perturbation theory which results in larger uncertainties \cite{Yerokhin2022}.

A more detailed discussion of the results including the transition frequencies, fine-structure splittings and the extraction of the nuclear charge radius is provided in the parallel publication \cite{Imgram23_PRL}. Here, we will focus on the different sources contributing to systematic uncertainties that have been investigated during the measurement campaign and are discussed in the next section.

\section{Uncertainties in quasi-simultaneous collinear and anticollinear laser spectroscopy}\label{sec:uncert}
A big advantage of quasi-simultaneous collinear and anticollinear laser spectroscopy is that most of the typical systematic frequency shifts from classical collinear spectroscopy cancel since they appear in both directions with opposite signs. The remaining uncertainty contributions are mainly caused by different conditions between collinear and anticollinear measurements, which will be discussed in the following sections.

\subsection{Ion start potential}
An important requirement is a stable starting potential. Otherwise, both laser beams probe different ion velocities and Eq.\,(\ref{eq:colacol}) is not valid, resulting in a systematic shift of the measured transition frequency $\nu_0$. Therefore, the time dependence of the ions' kinetic energy was investigated by observing the resonance line center position over a period of 80\,minutes. Drifts of more than 150\,MHz in one hour were observed corresponding to a potential drift of 0.87\,V/h. This is a typical value for the stability of high-voltage (HV) power supplies of the type used for the generation of the drift-tube potentials. In order to compensate this drift, the voltage applied to the central drift tube $U_\mathrm{A}$ was actively stabilized with a feedback-loop as explained in \cite{Müller20}: The HV potential is continuously measured with a precision high-voltage divider and small deviations from the nominal value are compensated through an additional low voltage in the range from 0 to 5\,V, generated with a digital-to-analog converter integrated on a data acquisition card. This reduced the drift by a factor of 5 from approximately 2.5\,MHz/min to 0.5\,MHz/min. The reason for the remaining drift is twofold: First, the electron current in the EBIS was constantly decreasing during the measurements. This increases the starting potential of the ions through the reduction of the negative space charge. According to the manufacturer, this behavior is untypical and is attributed to a damage of the electron cathode. Unfortunately, no replacement cathode was available for this measurement campaign. Secondly, it was observed that the ion beam position is also drifting with time and, thus, the angle between laser and ion beam changes, which leads to a variation of the spatial ion velocity distribution that is probed by the laser. Both effects influence the line-center position through the Doppler shift.

However, the remaining drift is largely compensated by performing a collinear-anticollinear (CA) measurement after an anticollinear-collinear (AC) measurement. For a linear line-center drift $\frac{\partial \nu_\mathrm{c/a}}{\partial t} = \mathrm{const.}$, the remaining systematic shift between the real transition frequency $\nu_0$ and the determined transition frequency from an averaged AC-CA pair measured with a constant time interval $\delta t \approx 5$\,min can be estimated as
\begin{equation*}
    \delta \nu_\mathrm{drift} = \nu_0 - \frac{\sqrt{\nu_\mathrm{a} (\nu_\mathrm{c}+\delta\nu_\mathrm{t})} + \sqrt{(\nu_\mathrm{c} + 2\delta\nu_\mathrm{t})(\nu_\mathrm{a} - 3\delta\nu_\mathrm{t})}}{2}
\end{equation*}
with 
\begin{equation*}
\delta\nu_\mathrm{t} = \frac{\partial \nu_\mathrm{c/a}}{\partial t} \cdot \delta t.
\end{equation*}
For realistic values of $\nu_0$, $\nu_\mathrm{c/a}$ and $\delta\nu_\mathrm{t} \approx 2.5$\,MHz, the systematic drift is estimated as $\delta \nu_\mathrm{drift} \approx 10$\,kHz, which is negligible in comparison to the targeted accuracy, other systematic sources, and the statistical uncertainty.

\subsection{Laser and ion beam alignment}
In classical collinear laser spectroscopy, the alignment of the laser and ion beam has a strong influence on the position of the line center. Already a small angle between both beams can introduce shifts of some MHz through the relativistic Doppler effect. For quasi-simultaneous collinear and anticollinear laser spectroscopy, this is strongly reduced as long as both laser beams are well aligned with respect to each other \cite{Krieger2017}. This behavior is expected to change significantly when a misalignment between the two laser beams is introduced. Therefore, a measurement series on different days with two different laser beam misalignments has been performed. The first configuration (a) was a horizontal crossing of the two laser beams with a separation of approximately 1\,mm in front of each laser entrance window to the beamline. This introduces an angle of $\arctan (2\,\mathrm{mm} / 5.2\,\mathrm{m}) \approx 0.38$\,mrad between the two beams and an effective horizontal displacement of the two beam profiles in the ODR of roughly 0.55\,mm. The second arrangement (b) was a vertical parallel displacement of the two beams of also roughly 0.55\,mm ($\approx$ the beam radius) so that the laser beams propagated parallel to each other without crossing. The configuration (b) has been tested with and without the velocity filter (VF) after the EBIS. Between those test measurements, reference measurements have been recorded where both beams were again superposed as good as possible.

The largest systematic shift of 8.6(7)\,MHz was observed with configuration (a). However, this shift cannot be solely explained by the introduced angle and must have a second origin. The explanation can be found in the spatial distribution of the ion velocity inside the beam which is not homogeneous. Although an ideally collimated ion beam is targeted, a residual divergence in the optical detection region and an additional small but finite horizontal energy dispersion due to the electrostatic switchyard is always present. Thus, two laterally displaced laser beams probe different velocity distributions with a different mean velocity $\beta$ even if they are perfectly parallel. This results in Doppler shifts for the two lasers that do not cancel in the geometric average of Eq.\,(\ref{eq:colacol}). Consequently, a larger systematic shift in $\nu_0$ is obtained. Any additional angle between the two laser beams can further increase or even decrease this shift depending on the probed distributions. This behavior has been reproduced qualitatively in numerical simulations of the ion beam trajectory with SIMION\,8.0. The spatial coordinates and the velocity vector of the individual ions in an analysis plane at the optical detection region were used to calculate the scattering rate of the individual ions for different laser frequencies of superposed collinear and anticollinear laser beams. The resulting simulated resonance spectra were analyzed for a variation of superimpositions of the laser beams and the ion beam. It became apparent, that a horizontal displacement of the two laser beams can indeed produce a systematic shift of $\nu_0$ with roughly the same size as observed in the experiment.

The vertical displacement configuration (b) exhibited a smaller shift than (a). The reason has also been found in the ion beam simulations: The divergence introduced by the switchyard is much smaller in the vertical than in the horizontal direction. However, it should be noted that the Wien filter additionally separates the ion velocities in the vertical direction. Therefore, a measurement series was performed without the velocity filter in operation. This did not lead to a change of the behavior and we conclude that the operation of the Wien filter does not introduce additional uncertainties.

The largest shift of $8.6(7)$\,MHz is produced with setting (a), where the laser displacement was about $1\,\mathrm{mm} / 0.2\,\mathrm{mm} = 5$ times larger than under usual experimental conditions. Thus, a systematic uncertainty contribution for the laser beam alignment due to the spatial velocity dispersion (VD) in the beam is estimated to $\Delta \nu_\mathrm{VD} = 8.6\,\mathrm{MHz} / 5 \approx 1.7$\,MHz. This range indeed covers all reference measurements taken during the investigations and is only slightly larger than twice their $1\sigma$-standard deviation of 0.8\,MHz.

\subsection{Photon recoil}
A photon carries a momentum $p_\gamma = h\nu/c$ additional to its energy $E = h \nu$. Energy and momentum are conserved during absorption and emission processes in an ion. This means that a part of the photon energy is added to the kinetic energy of the ion instead of the internal transition energy $\Delta E = h \nu_0$ which results in slightly higher measured resonance frequencies than $\nu_0$. Therefore, the determined rest-frame frequency must be corrected by the recoil frequency $\delta \nu_\mathrm{rec} = h \nu_0^2 / (2mc^2)$ for comparison with \textit{ab initio} calculations that calculate the difference of level energies. This is done during the analysis process and does not introduce an additional uncertainty.

Another consequence of the absorption of a photon is the change of the ions momentum by the size of the photon momentum. Near resonance interaction with several excitation-emission cycles, leads on average to an acceleration of the ion in the collinear setup and a deceleration in the anticollinear setup. In both cases the induced velocity change requires a higher laser frequency in the laboratory frame for  the next excitation. Thus, a frequency shift towards a higher rest-frame frequency could in principle occur if more than one excitation takes place in the ODR. The mean number of scattered photons $\overline{n}_\mathrm{sc}$ in resonance is estimated to be 1 to 4 for laser intensities between half of the saturation intensity and the saturation intensity. The velocity change per excitation is $\delta \upsilon = h \nu_\mathrm{c/a}/(m c) \approx 0.146$\,m/s, resulting in a center frequency shift towards higher laser frequencies of $\partial \nu_\mathrm{c/a}/\partial \upsilon \cdot \delta \upsilon \approx 0.65$\,MHz per excitation and emission cycle. Hence, theoretically, an increase of the observed rest-frame frequency is expected for increasing laser intensity.

However, no significant shift outside of the main uncertainty $\Delta \nu_\mathrm{VD}$ was observed in a measurement series with different laser intensities. A probable reason is the remaining divergence of the ion beam, preventing single ions to be excited over the full length of the ODR. This reduces the mean number of scattering events of an individual ion which was calculated under the assumption of an ion in resonance with the laser over the full length of the ODR. We note that the majority of measurements were performed with half of the saturation intensity which makes systematic shifts from photon recoil negligible with respect to the current statistical uncertainty and the main systematic uncertainty $\Delta \nu_\mathrm{VD}$. Nevertheless, this effect needs to be investigated in more detail once the main systematic uncertainty $\Delta \nu_\mathrm{VD}$ has been reduced.

\subsection{Other uncertainties}
A possible misalignment of the ion beam with respect to the two laser beams was investigated and found to contribute insignificantly to the uncertainty in accordance with previous investigations \cite{Krieger2017}. Similarly, the correction of a small mismatch ($\delta U < 0.2$\,V) in the scan voltage applied at the optical detection region between the collinear and the anticollinear resonance centers according to \cite{Krieger2017}
\begin{equation}
    \nu_0 = \sqrt{\nu_\mathrm{a} \cdot \left(\nu_\mathrm{c} - \frac{\partial \nu}{\partial U}\cdot \delta U \right )} 
\end{equation}
introduces an uncertainty of less than 70\,kHz, which is negligible compared to the beam alignment uncertainty.
We also investigated the influence of the laser light polarization since circularly polarized light can systematically shift the resonance frequency due to the Zeeman splitting in a residual magnetic field. Since the Zeeman splitting is not resolved, we addressed simultaneously all possible $\Delta m=0$ Zeeman transitions with the linearly polarized laser light. This broadens the resonance slightly, but does not shift the center frequency. We investigated the influence of impurities of the laser polarization to the extracted rest-frame transition frequency. Such an impurity can be caused, for example, by stress-induced birefringence in the viewports of the beamline. Measurements were performed using circularly and elliptically polarized light but a shift outside the range of the always present laser-alignment uncertainty was not observed, even with purely circular polarization. Therefore, this effect was neglected at the currently achievable level of accuracy.

\begin{table}[htb]
    \centering
    \caption{All investigated systematic uncertainties in MHz. The main systematic uncertainty originates from the alignment of the two laser beams. Details can be found in the text.}
    \begin{minipage}{0.28\textwidth}
    \begin{ruledtabular}
        \begin{tabular}{cd}
        {Uncertainty} & \\\hline
        Line shape & <0.01\\
        Start potential & 0.01\\
        Beam alignment & 1.7\\
        Photon recoil & <0.1\\
        Scan voltage & <0.07\\
        Laser polarization & 0\\\hline
        Total & 1.7
        \end{tabular}
        \end{ruledtabular}
    \label{tab:uncertainties}
    \end{minipage}
\end{table}

In summary, the main systematic uncertainty is given by $\Delta \nu_\mathrm{VD} = 1.7$\,MHz which emerges from a possible residual misalignment of the two laser beams in combination with the ion beam divergence. It currently limits the achievable precision, but it still represents an improvement of more than three orders of magnitude compared to the previous experimental values \cite{Ozawa01}.

\section{Outlook}
The next isotope of interest is $^{13}$C due to its nuclear spin which introduces the hyperfine structure to the optical spectrum. This will challenge the experiment as well as the theory. However, when the \PJ\ transition frequencies in $^{13}$C$^{4+}$ can be extracted with similar precision as in this work, the nuclear mean-square charge radius of $^{13}$C can be determined from the optical isotope shift with a precision only limited by the muonic x-ray spectroscopy result for $^{12}$C \cite{Ruckstuhl84}. The conventional approach using mass-shift calculations in the two electron system will be applied as it has been done before to investigate the short-lived isotopes $^6$He \cite{Wang2004}, and $^8$He \cite{Mueller2008}.  

After carbon, we will tackle the two naturally abundant boron isotopes $^{10,11}$B. Here, in addition to the He-like charge state, we will also investigate the $2\,^2$S$_{1/2} \rightarrow 2\,^2$P$_{1/2}$ transition (206\,nm) in Li-like B$^{2+}$. Besides the all-optical charge radius determination, the isotope shift measurement in both charge states will enable a thorough comparison between mass-shift calculations in light two-, three- and five-electron systems \cite{Maaß2019}. In order to produce B$^{2+,3+}$ in an EBIS, a volatile organic compound (trimethyl borate BO$_3$C$_3$H$_9$), which has a high vapour pressure, will be fed into the EBIS. First tests have already been performed at GSI, Darmstadt \cite{Mohr2023}.

Since for beryllium no volatile organic compound exists, the investigation of Be$^{2+}$ requires to first produce Be$^+$ in a different ion source and then inject it into the EBIS for charge breeding. 

\section{Summary}
We successfully performed high-precision collinear laser spectroscopy in the \PJ\ transitions of He-like $^{12}$C$^{4+}$. In order to enable collinear laser spectroscopy on highly charged ions, we upgraded the COALA beamline at TU Darmstadt with a new electron beam ion source including a Wien-filter for charge-to-mass separation. Additionally, we implemented a new switchyard which allows us to operate up to three ion sources installed permanently to quickly switch between different ion species. This allows us to externally feed ions from a different source into the EBIS for charge breeding. In order to optimize signal rates and spectral resolution of collinear laser spectroscopy with $^{12}$C$^{4+}$, we investigated the bunched beam and continuous beam mode. We found that the continuous beam mode yields the narrowest spectral linewidth with a FWHM of 170\,MHz originating from the 1-V energy width of the EBIS. Furthermore, we investigated several systematic effects influencing the determination of transition frequencies in frequency-comb-referenced quasi-simultaneous collinear and anticollinear laser spectroscopy and found that the largest contribution originates from the remaining ion beam divergence in combination with a slight misalignment of the two laser beams resulting in a systematic uncertainty of 1.7\,MHz. This led to an improvement of the \PJ\ transition frequencies of more than three orders of magnitude compared to previous experimental results and tested recent QED calculations of He-like carbon with high precision. Additionally, an all-optical nuclear charge radius of $^{12}$C was extracted which is published in \cite{Imgram23_PRL}.



%
%

%

\begin{acknowledgments}
We thank J.~Krämer for his contributions in the early stage of the project and K.~Pachucki \& V.~Yerokhin for many fruitful discussions. We acknowledge funding by the Deutsche Forschungsgemeinschaft (DFG) Project No. 279384907 SFB 1245, as well as under Grant INST No.\ 163/392-1 FUGG, and from the German Federal Ministry for Education and Research (BMBF) under Contract No. 05P21RDFN1. P.I. and P.M. acknowledge support from HGS-HIRE.
\end{acknowledgments}

%

\end{document}